\documentclass[nofootinbib,superscriptaddress,reprint,onecolumn,notitlepage,12pt]{revtex4-1}
\usepackage{hyperref}
\usepackage{graphicx,natbib}

\begin{document}

\title{Structure in galaxy clusters}

\author{G.B. Lima~Neto}
\affiliation{IAG/University of S\~ao Paulo, Brazil}
\author{T.F. Lagan\'a}
\affiliation{NAT/Unicsul, S\~ao Paulo, Brazil}
\author{F. Andrade-Santos}
\affiliation{Harvard-Smithsonian CfA, Cambridge, MA, USA}
\author{R.E.G. Machado}
\affiliation{IAG/University of S\~ao Paulo, Brazil}

\begin{abstract}
We will discuss here how structures observed in clusters of galaxies can provide us insight on the formation and evolution of these objects. We will focus primarily on X-ray observations and  results from hydrodynamical $N$-body simulations. This paper is based on a talk given at the School of Cosmology \textit{Jos\'e Pl\'{i}nio Baptista} -- ``Cosmological perturbations and Structure Formation'' in Ubu/ES, Brazil.
\end{abstract}

\maketitle

\section{Introduction: Cluster formation}

\begin{figure}[htb]
\centering
\includegraphics*[width=16cm]{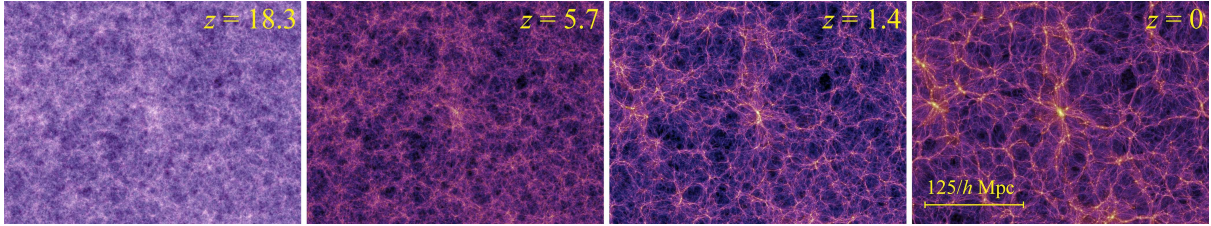}
\caption[]{Snapshots from the Millennium Simulation, run at the Max Planck Society's Supercomputing Centre in Garching, Germany, \citep{Springel05} showing the time evolution of a slice of the Universe. Large haloes are formed at the intersection of cosmic dark matter filaments by accreting mass. This figure were made with the images available at
\texttt{www.mpa-garching.mpg.de/galform/virgo/millennium/}. }
\label{fig:Springel_LCDMsma}
\end{figure}

\begin{figure}[htb]
\centering
\includegraphics*[width=5.0cm]{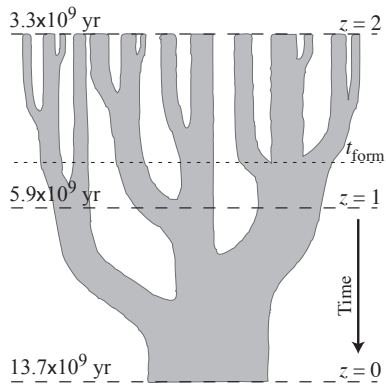}
\caption[]{ Merging tree showing the events leading to the formation of a massive dark matter halo at $z = 0$. $t_{\rm form}$ corresponds to the instant that half of the present halo mass is already assembled.  Figure adapted from \citet{Lacey93}.}
\label{fig:MergerTree}
\end{figure}

In the current accepted cosmological scenario, where the matter content of the Universe is dominated by ``cold'' collisionless dark matter, clusters of galaxies are the last collapsed structures to form. $N$-body cosmological simulations \citep[e.g.,][]{Springel05,Klypin11,Alimi2012} show clearly that, in a cold dark matter dominated Universe, halos of $10^{14}$ to a few $10^{15} M_\odot$ form at the nodes or intersections of large scale cosmic filaments. Large halos are built hierarchically by accreting smaller objects, as shown in cosmological simulations (Fig.~\ref{fig:Springel_LCDMsma}) and represented as a ``merging tree'' (Fig.~\ref{fig:MergerTree}). Therefore, even though clusters are usually taken as already formed structures, we should expect matter infall and accretion along the cosmic filaments, which may still impact cluster evolution. 

The hierarchical formation of collapsed structures will determine the halo mass function through the density fluctuation power spectrum. The first cosmological analysis of the mass function in a  dark matter dominated Universe was made by \citet{Press74}, where they assumed that primordial density fluctuations were Gaussian, and that regions at any given time which exceeds a certain density constant are collapsed,.

Even with its shortcomings (an \textit{ad hoc} factor 2 that had to be inserted in the mass function), their mass function was remarkably accurate compared to $N$-body simulations. During the last 15 years, the halo mass function was modified using the results of cosmological simulations \citep[e.g.][]{Jenkins01, Tinker08}.  

\begin{figure}[htb]
\centering
\includegraphics[width=15cm]{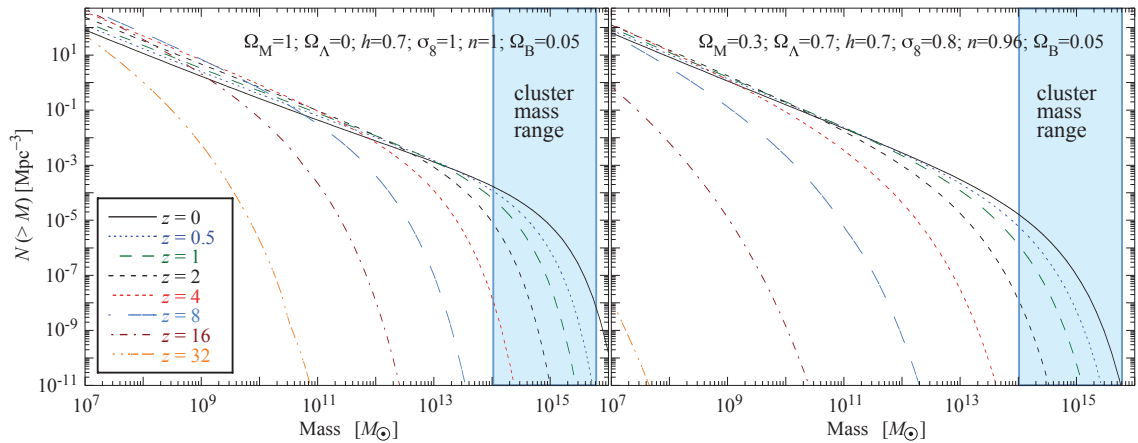}
\caption[]{Halo mass function as a function of mass at different redshifts. \textsf{Left}: Mass function for an Einstein-de Sitter Universe (flat, $\Omega_M =1$). \textsf{Right}: Mass function for a standard $\Lambda$CDM scenario. The curves are computed using the Press-Schechter mass function and an approximate power spectrum similar to the one given by \citet{Tegmark04}.}
\label{fig:Press_Schechter_N_M}
\end{figure}

The mass function is an important cosmological probe, in particular for the mean mass density of the Universe, $\Omega_M$, and the amplitude of the density fluctuation power spectrum. Fig.~\ref{fig:Press_Schechter_N_M} illustrates how the mass function is different for different cosmological parameters. Notice that the abundance of rich clusters with $M > 10^{14} M_\odot$ gives strong constraints for the halo mass function.

\section{Clusters of galaxies}

Clusters of galaxies are the largest collapsed structures in (quasi) equilibrium in the Universe. Rich clusters have thousands of galaxies but, since they are relatively rare only about 5--10\% of galaxies live inside clusters \citep[see][for a review]{Bahcall99}.

It is usual to refer to collapsed structures with mass greater than about $10^{14} M_\odot$ as clusters, while for objects with masses between $\sim 10^{13}$ and $10^{14}$ we call them groups. The typical radii of clusters are 1--3~Mpc (a value close to the Abell radius, $1.5 h^{-1}$Mpc or $R_{A} = (1.72/z)$~arcmin). The mean density is $\sim 200$ times the critical density of the Universe at the cluster redshift.

Galaxy clusters are also characterized by their galaxy population, largely dominated by early-type galaxies (elliptical and lenticular), contrary to the field where spiral galaxies are more common.

The galaxies in clusters correspond only to $\sim 2$--5\% of the total mass. Most of the baryonic matter is in the form of a hot diffuse plasma, the intracluster gas, amounting to about 12--15\% of the total mass, i.e., there is roughly 6 times more baryons in the gas than in all galaxies together. However, most of the mass, about 85\% is in the form of dark matter.

The dark component cannot be directly observed; its presence is inferred by gravitational effects. Under the hypothesis that clusters are in equilibrium, the total mass may be obtained using either the galaxies as tracer (by the virial theorem) or the intracluster gas, by hydrostatic equilibrium, which results in the well know formula for the total mass inside the radius $r$:
\begin{equation}
 \frac{1}{\rho} \vec{\nabla} P = -\vec{\nabla} \Phi \quad \Rightarrow \quad  M(r) = - \frac{kT}{G \, \mu\, m_{\rm H}}\, r \, 
  \left(\frac{d \ln \rho}{d \ln r} + \frac{d \ln T}{d \ln r} 
  \right)\, ,
\end{equation}
where we have assumed spherical symmetry.

A third evidence of dark matter comes without the equilibrium hypothesis from gravitational lens effect -- the cluster mass acts as a gravitational lens on the background galaxies and the total projected mass along the line-of-sight may be estimated \citep[see][]{Bartelmann01, Kneib11}

\subsection{The intracluster gas}

The intracluster gas has typical temperature ranging from $10^7$ to $10^8$~K, that are generally given in the literature in energy units: $2 < kT < 12$~keV, approximately. It is a very rarefied plasma, with typical central number densities of $n_0 \approx 10^{-2}$--$10^{-3}$~cm$^{-3}$. Such a plasma is a strong X-ray source emitting through thermal bremsstrahlung, i.e. free--free scattering process of electrons by ions \citep[see, e.g.,][]{Sarazin88, Bohringer10}.

\begin{figure}[htb]
\centering
\includegraphics[width=13cm]{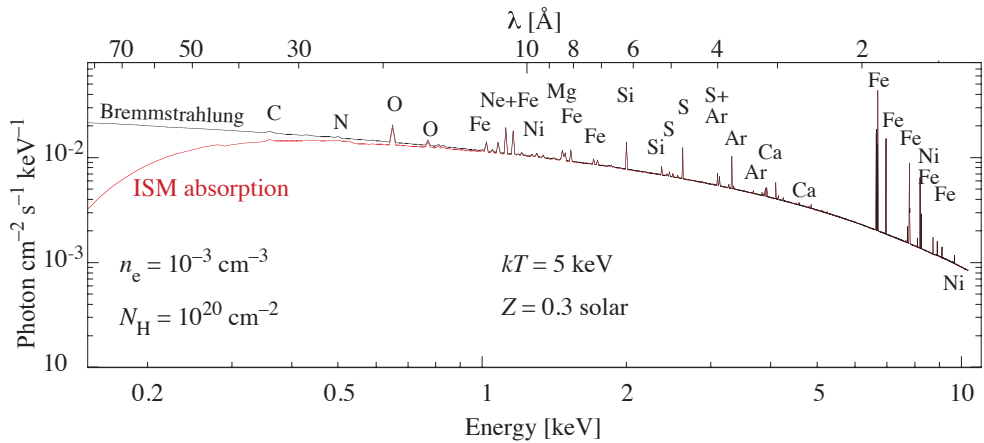}
\caption[]{Theoretical emission from the intracluster gas, bremsstrahlung continuum plus emission lines. The red curve corresponds to the absorbed spectrum, mainly from the interstellar HI from our Galaxy.}
\label{fig:R-S_emissaoX}
\end{figure}

The intracluster plasma is metal enriched by processed gas in galaxies, having a typical mean metallicity of a third of the solar metallicity. The heavier elements are responsible for producing the recombination lines observed in the X-ray spectrum of clusters. The expected emission from a 5~keV cluster is shown in Fig.~\ref{fig:R-S_emissaoX}.

The gas looses energy by bremsstrahlung emission, which is approximately proportional to $n^2 T^{1/2}$, so we can define a cooling-time as:
\begin{equation}
t_{\rm cool} \approx \frac{E}{d E/d t} \quad \Rightarrow \quad
9.3 \times 10^9 \frac{(k T_{\rm keV})^{1/2}}{n_3} \mbox{years} \, ,
\end{equation}
where $kT$ is given in keV and $n_3$ is the gas density in units of $10^{-3}$~cm$^{-3}$, and $E$ and $dE/dt$ are the thermal energy and the cooling ratio of the gas.

The relatively short cooling times at the core of clusters have led to the scenario where cool gas would flow to the cluster center at rates up to $\sim 1000 M_\odot/$yr, a phenomenon know as cooling-flow \citep{Fabian84, Fabian94}. The lack of great amounts of cool gas, no evidence of ongoing star formation, and, observations of X-ray spectra led to the actual scenario where the gas is heated and the cooling-flow is greatly reduced. The most popular hypothesis is that the activity of the central AGN is regulated by the infall of gas producing a feedback mechanism that keeps the gas temperature at roughly 1/3 of the virial temperature value \citep[e.g.,][]{McNamara07}.

\begin{figure}[htb]
\centering
\includegraphics[width=14.5cm]{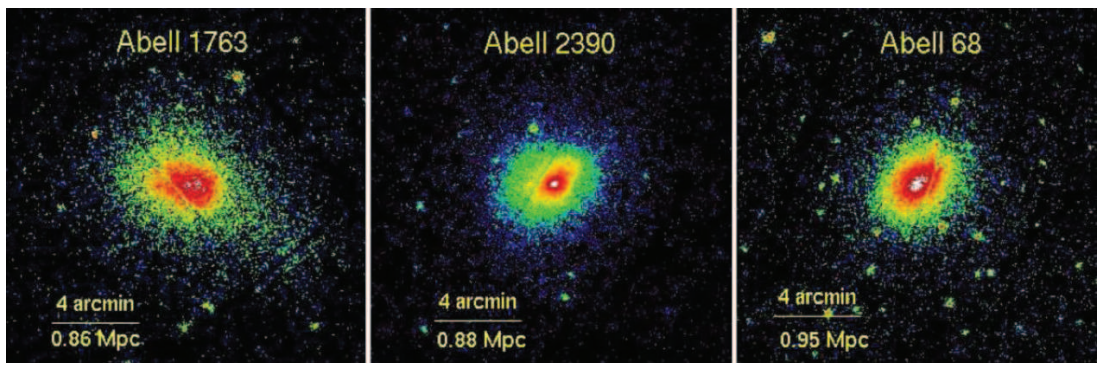}
\caption[]{Example of three clusters observed with XMM-\textit{Newton}. The images show the X-ray surface brightness. Even for relaxed clusters it is possible to see that they are not perfectly regular (i.e., axial symmetric), indicating some previous dynamical activity.} 
\label{fig:3amasX}
\end{figure}

Still, due to its short relaxation time, the intracluster gas is an important probe of cluster dynamics. The X-ray surface brightness of clusters (Fig.~\ref{fig:3amasX}), which is the projection of $n^2$ on the plane of the sky, reveals their dynamical state, the presence of substructures and on-going collisions. Temperature maps are even more revealing, showing the presence of shocks and cold fronts moving through clusters.

\begin{figure}[htb]
\centering
\includegraphics[width=8cm]{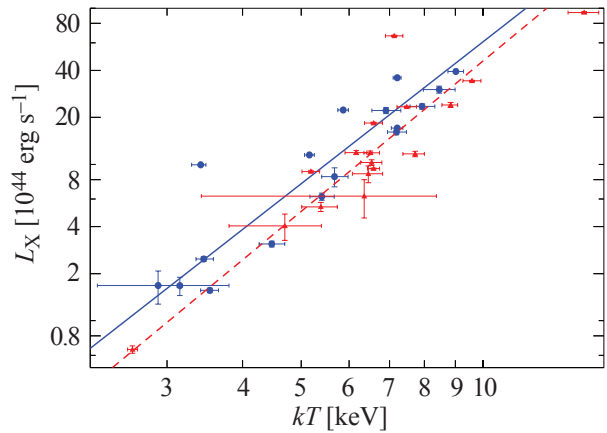}
\caption[]{$L_X$--$T_X$ relation fitted separately for clusters showing a high level of substructures (red dashed line) and a low level of substructures (blue continuous line). For a given luminosity, substructured clusters are hotter then more relaxed clusters. Figure taken from \citet{Andrade12}.}
\label{fig:TxLx}
\end{figure}

With very deep imaging one can perform detailed morphological analysis, specially of nearby clusters, that shows a wealth of substructures: bubbles (cavities), filaments, and even acoustic waves. Some of them are driven by non-gravitational process linked to the central AGN activity. 
These structures impact on the global cluster properties and affect their scaling relations (such as $T_X$--$L_X$) currently used, for example, for cosmological studies. Figure~\ref{fig:TxLx} shows an example of the kind of bias that may be introduced when substructures are ignored when selecting a sample. This motivates the detailed study of individual clusters both from the morphological point of view and dynamically, helped by $N$-body simulations.

\subsection{Search for substructures in clusters}

Motivated by the dynamical information we can infer from the analysis of irregular morphology in clusters, many authors have searched and quantified substructures in clusters. Focusing on X-ray observations, the first systematic study was done by \citet{Jones84} based on \textit{Einstein} data.

Observation of the surface brightness allows us to derive quantitatively the degree of substructure of clusters. Various tests were proposed in the literature: centroid and ellipticity level variation \citep{Mohr95}, moments of the expansion in Fourier series \citep{Buote95}, power-ratio method \citep{Jeltema05}, ratio of fluxes between residual and original X-ray images \citep{Andrade12}, among others.

\begin{figure}[htb]
  \centering
  \includegraphics[width=8.5cm]{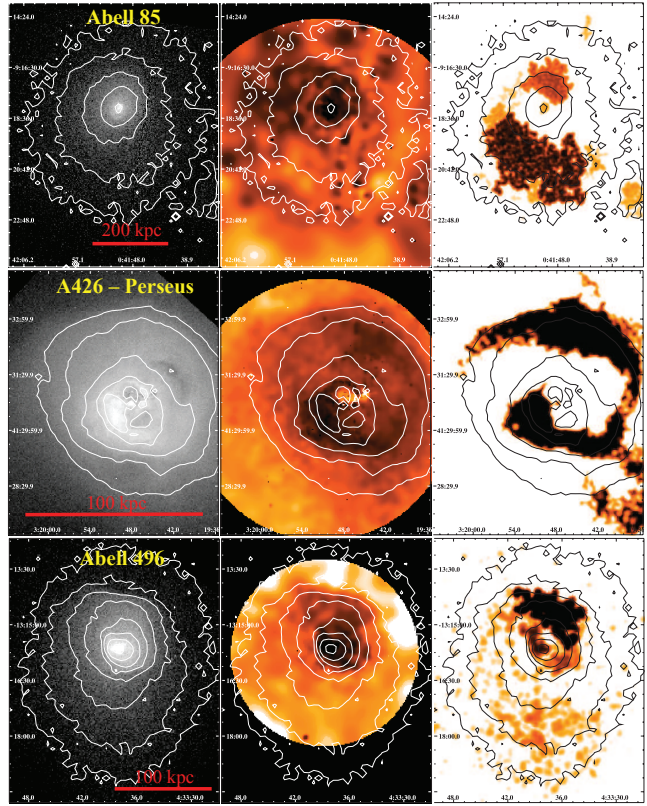}
 \caption[]{From left to right: surface brightness image, hardness ratio (temperature) map, and residual substructure distribution, showing the most prominent arm substructure (from \citet{Lagana10}), seem in both temperature and substructure maps.}
  \label{fig:simple}
\end{figure}

\citet{Lagana10} chose a different approach, using a qualitative analysis based on the residual image after subtraction of a smooth elliptically symmetrical model. Selecting a sample of 15 nearby clusters, $z < 0.06$, i.e., within $250 h_{70}^{-1}\,$Mpc from us, observed by \textit{Chandra} for more than 30~ks, they examined the surface brightness and temperature map morphologies.  Figure~\ref{fig:simple} shows the large scale substructures in three of these clusters. The most striking feature is the spiral arm structure.

In half of their sample a spiral-like arm substructure is observed, all of them in cool-core clusters. Such a substructure is more easily seen in the residual map than in the hardness ratio map.  All clusters show, in some degree, substructures (not necessarily an arm structure).

\begin{figure}[htb]
\centering
\includegraphics[width=13cm]{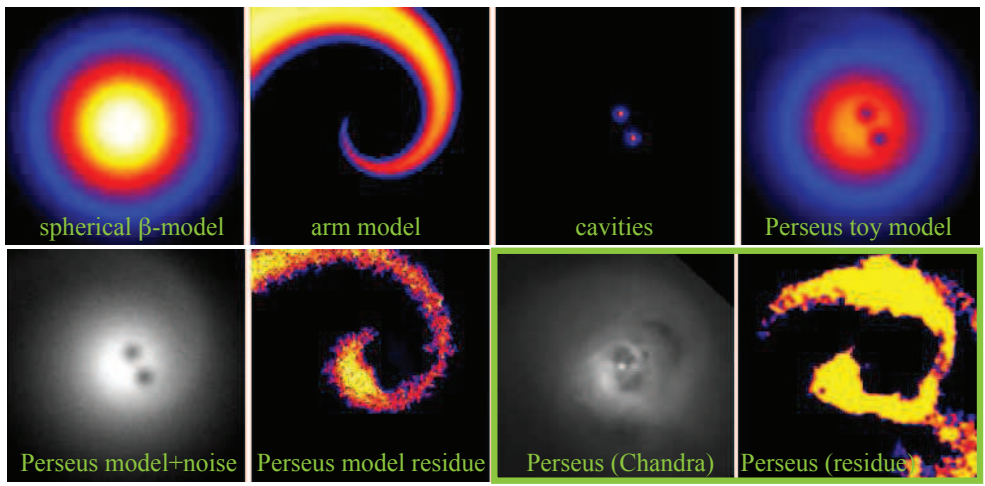}
\caption[]{Test of detectability of the spiral feature (arm) in clusters. \textsf{Top}: construction of a toy model of the Perseus X-ray surface brightness. \textsf{Bottom}: application of the same method for the toy model and \textit{Chandra} Perseus data. Figure from \citet{Lagana10}}
\label{fig:armModelPerseus}
\end{figure}

In order to verify if the arm feature was not an artifact of the image processing algorithm, \citet{Andrade12} has performed several tests with simulated cluster images (cf. an example in Fig.~\ref{fig:armModelPerseus}). Gas hydrodynamics is essential to be well understood if we want to use X-ray observables as mass estimators.

The arm substructures have a scale of hundreds of kiloparsecs, larger than the region where cavities are produced by the AGN activity. The mechanical work, $P\,dV$, from the AGN may not be enough to push the gas and generate these observed arms. The work required to form an arm with $\sim 200$~kpc is roughly $\sim 10^{61}$~erg or $5 \times 10^{45}$~erg~s$^{-1}$ for $10^8$ years. Minor collisions (that may lead to a merger) may provide an alternative source of energy through tidal forces, by ``shaking'' the cool-core. A small cluster/rich group with $10^{14} M_\odot$ moving with a modest velocity of 500~km~s$^{-1}$ has $5\times 10^{62}$~erg of kinetic energy; only a small fraction of this energy would be enough to generate a large scale spiral arm on the intracluster gas. 

\citet{Ascasibar06, ZuHone11} have shown with $N$-body simulations that off-center collisions of a cluster with a galaxy group -- mass ratio between 1/2 to 1/10 -- may slosh the cool central gas and produce the arms we observe. Notice that this mechanism is efficient to move the cool gas from the center but it is not a heating mechanism to stop or regulate the cooling-flow.

\subsubsection{Cluster--AGN connection}

Even though cluster AGNs may not be powerful enough to generate large scale substructures, the radio jets -- relativistic particles ejected from near the central massive black hole -- interacts with the intracluster gas and have important consequences. The favored mechanism to suppress the cooling-flow is AGN feedback \citep[e.g., review by][]{McNamara07, Fabian12}.

\begin{figure}[htb]
\centering
\includegraphics[width=7.5cm]{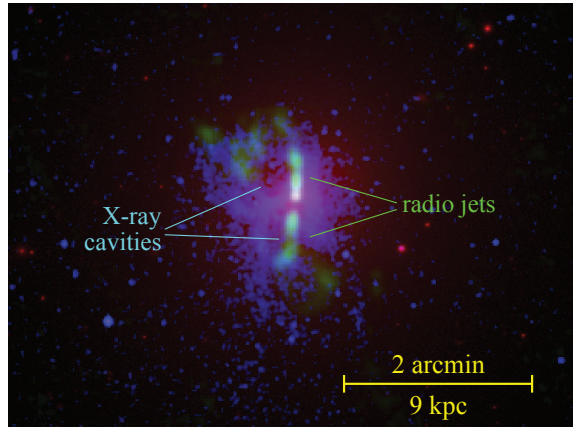}
\caption[]{Composite image of M84. Red: SDSS $r$ band image; Green: VLA 1.4 First Survey; Blue: Chandra 0.3--7.0 keV band. The X-ray ``cavities'' and radio jets are indicated.}
\label{fig:M84_SDSSr_VLA_ACIS}
\end{figure}

Most, if not all, Brightest Cluster Galaxies (BCGs) have radio emission and some present strong jets (e.g., Cygnus~A, Hydra~A, M~84, NGC~1275 [Perseus~A]) that blow ``cavities'' (also called bubbles or holes) in the intracluster gas (see, e.g., Fig.~\ref{fig:M84_SDSSr_VLA_ACIS}), observed in high resolution X-ray images. The evidence for this lies in the spatial coincidence between the observed synchrotron radio emission and the regions where there is a decrease in X-ray flux. Those cavities are also buoyant and probably migrate outwards from the center, towards lower density regions.

Very deep imaging of the Perseus Cluster \citep{Fabian11} confirms earlier finding that the AGN activity through their jets produces acoustic (pressure) waves that propagate outwards throughout the intracluster plasma. A challenging task would be to detect these waves in other clusters. Therefore, it seems plausible that the AGN energy is converted to mechanical and thermal energy, heating the gas in a feedback process.

\section{Cluster collisions}

Massive clusters are built by way of major cluster collisions leading to a hierarchical merging process. Not only does the study of merging clusters provide a way to understand large scale structure formation in the Universe, but it also helps us understand galaxy evolution, the interplay between the intracluster medium and galaxies, and the nature of dark matter.

One way to approach this problem is to follow the growth of clusters with cosmological simulations, with a minimum of \textit{ad hoc} assumptions. This requires a great amount of computational power, that is still somewhat limited by mass/spatial resolution in particular if the simulation is hydrodynamical -- usually only collisionless dark matter is simulated \citep[see, e.g.,][]{Kuhlen12}.

Another option is to simulate individual systems with self-consistent $N$-body hydrodynamical simulations, sometimes even with sub-grid physics, i.e., star formation, feedback, cooling, metal enrichment, and other physical process usually called ``gastrophysics'' by the community, and are dealt with semi-analytical models \citep[e.g.][]{White91,Lacey91,Somerville08} The main setback here is to ignore the continuous matter accretion in clusters in a self-consistent way, focusing only in a given event, a cluster-cluster collision.

\subsection{The Bullet Cluster 1E~0657-558}

The most studied colliding cluster today is doubtless 1E~0657-558, also known as the \textit{Bullet Cluster} \citep{Markevitch02}. This cluster, located at redshift $z = 0.296$, acquired its nickname because of a prominent bow shock observed with high resolution X-ray imaging, suggesting that a smaller cluster is passing through a lager cluster at high speed: the bow shock appears as a consequence of the gas collision at Mach number $\sim 3$ almost on the plane of the sky.

Deep optical imaging shows two mass concentrations detected from their weak lensing signal, spatially coincident with the galaxy distribution of both colliding clusters (the ``bullet'' and the main cluster), but displaced from the two gas concentrations (detected in X-rays). $N$-body hydrodynamical simulations \citep[e.g.,][]{Springel07, Mastropietro08} confirm the collision scenario of the Bullet Cluster, where the relative velocity of the clusters halos is $\sim 2700$~km~s$^{-1}$ (actually smaller than the shock speed), and reproduce the observed separation of the bulk of mass from the intracluster gas. The displacement of the intracluster plasma with respect to the dominating halo dark matter is due to the gas dissipative nature and provides strong constraints for the hypothetical dark matter particles cross section \citep[in the simplest scenario, dark matter is collisionless,][]{Clowe06}.

There are a few other clusters that were recently discovered to have the gas component dissociated from the dark matter, for instance, Abell 2744 \citep{Merten11}, Abell 1758 \citep{Ragozzine12}, and DLSCL J0916.2 \citep{Dawson13}.

\subsection{Abell 3376 -- a nearby bullet}

\begin{figure}[!htb]
\centering
\includegraphics[width=12.5cm]{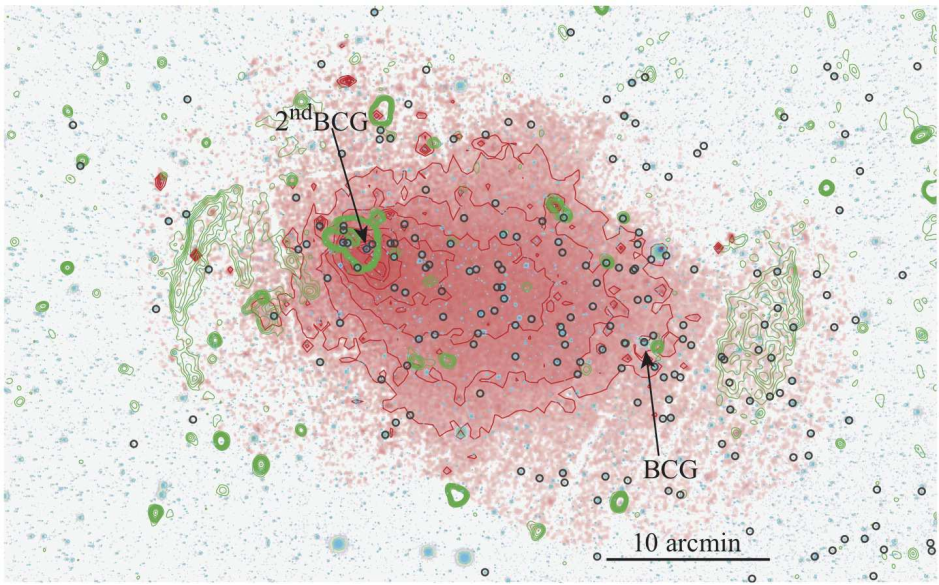}
\caption[]{XMM-\textit{Newton} image of Abell 3376 in the [0.3--8.0] keV band (red shades and contours) with radio VLA 1.4 GHz contours (green) and Blanco/Mosaic II B-band image (blue). The black circles represent galaxies in the $0.036 < z < 0.056$ redshift range. The two brightest galaxies are indicated.}
\label{fig:A3376_VLA_XMM_Blanco}
\end{figure}

The cluster Abell 3376 is a nearby cluster, $z = 0.046$, that shows signs of having passed recently through a major collision \citep{Bagchi06}. The XMM-\textit{Newton} X-ray surface brightness map shows a cometary-like morphology, similar to 1E~0657-586. Moreover, the brightest cluster galaxy (BCG) is displaced from the central X-ray emission peak, the gas temperature map shows alternating hot and cold gas, and deep VLA observation shows a huge ($\sim 2 \times 1.6$~Mpc on the plane of the sky) diffuse synchrotron radio emission that suggests a structure in an ellipsoidal distribution around the cluster (Fig.~\ref{fig:A3376_VLA_XMM_Blanco}). This radio emission is possibly generated by electrons accelerated by Fermi process due to the shock of cluster collision. Analyses of the galaxy distribution and dynamics \citep{Escalera94,Ramella07,Durret13} show that Abell~3376 has an elongated, substructured distribution along the putative collision path.

\begin{figure}[!htb]
\centering
\includegraphics[width=13.5cm]{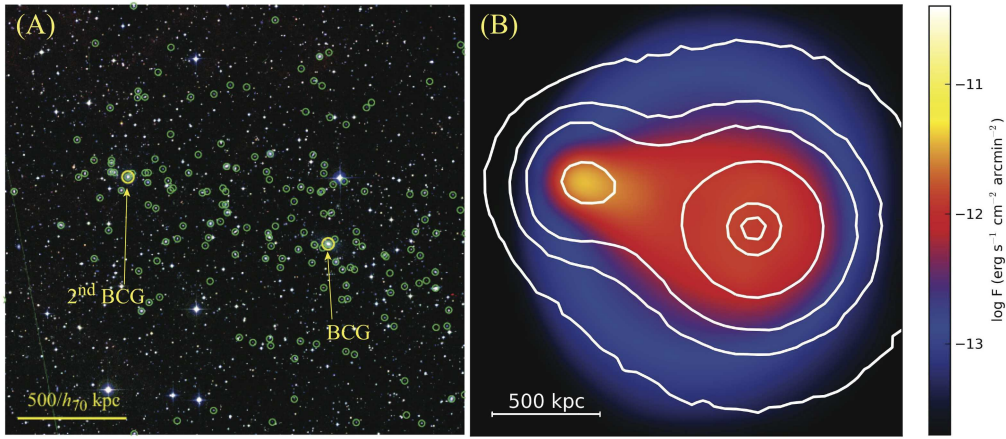}
\caption[]{(A) DSS optical image in which the circles mark positions of the galaxies at the cluster redshift. The first BCG (right) and the second BCG (left) are highlighted. \textbf{(B)} Simulated X-ray surface brightness after the cluster collision, where the
white contour lines show the total projected mass from \citet{Machado13} best model.}
\label{fig:Fig2_A3376_DM_galaxydist}
\end{figure}

Recently, \citet{Machado13} carried out a large set of high-resolution adiabatic hydrodynamical $N$-body simulations, in order to verify and model the collision event of Abell 3376. They were able to propose a specific scenario for the collision: the most successful model -- a 1:6 mass ration, head-on collision at Mach $\sim 3.5$ viewed with a $40^{\circ}$ with respect to the plane of the sky -- accounts for several of the features observed in the cluster (gas morphology, temperature, virial mass, total X-ray luminosity). Interestingly, a testable theoretical prediction emerges from this scenario, concerning the distribution of dark matter. In the resulting configuration, the dark matter (and therefore the bulk of the total mass) is predicted to be concentrated around two density peaks where the cluster brightest galaxy (BCG) and the second BCG are located (Fig.~\ref{fig:Fig2_A3376_DM_galaxydist}). 

It is intriguing to notice that, contrary to the Bullet Cluster, the \textit{simulated} mass distribution seems to follow the \textit{simulated} gas distribution. The \textit{real} mass distribution of Abell 3376 is, as of yet, unknown. If the mass map approximately matches the simulation results, this would lend more weight to our scenario, and might even set tighter constraints. If, on the other hand, the mass map reveals some complicated or unexpected dark matter structure, then the current models will be shown to be insufficient, and alternative models will have to be explored. Hence, mapping the total mass distribution through gravitational weak lensing effect on background galaxies is an important step for our understanding of this system.

\section{Concluding remarks}

Galaxy clusters are fundamental cosmological probes and giant astrophysical laboratories. In order to fully use these objects for precision cosmology and astrophysics, we must well grasp the physical process occurring with them. We briefly explored here the possibilities open by the X-ray astronomy and numerical simulations of cluster interactions, discussing a few recent results on substructure quantitative determination, AGN feedback, the consequences of off-center collisions and sloshing, and head-on supersonic collisions generating bow shocks.

\paragraph*{Acknowledgments:} {\small 
GBLN thanks the Brazilian agencies CNPq and CAPES for financial support and the hospitality during the \textit{Jos\'e Pl\'{i}nio Baptista} School. TFL and REGM thank FAPESP (fellowships 2012/00578-0 and 2010/12277-9, respectively) We also acknowledge enjoyable and fruitful discussions with Florence Durret and Melville Ulmer.}

\end{document}